\documentclass[10pt]{blois07}
\usepackage{graphicx}
\usepackage{amsmath,amssymb}
\usepackage{cite,./mcite}

\begin{document}
\title{Fitting DVCS at NLO and beyond}
\author{
 K. Kumeri\v{c}ki$^1$,
 D. M\"{u}ller$^2$,
 K. Passek-Kumeri\v{c}ki$^3$\protect\footnote{%
\hspace*{1ex}Based on talks presented by the authors
at
{\it Elastic and Diffractive Scattering 2007}, DESY Hamburg, Germany;
{\it Exclusive Reactions at High Momentum Transfer}, JLab Newport News, US;
{\it The 6th Circum-Pan-Pacific Symposium on High Energy Spin Physics}, UBC Vancouver, Canada;
{\it XII  Workshop on High Energy Spin Physics}, JINR Dubna, Russia;
and {\it New Trends in High-Energy PHYSICS}, Yalta, Ukraine. } }
\institute{
$^1$Department of Physics, Faculty of Science, University of Zagreb, Croatia\\
$^2$Institut f\"ur Theoretische Physik II, Ruhr-Universit\"at Bochum, Germany\\
$^3$Theoretical Physics Division, Rudjer Bo{\v s}kovi{\'c} Institute, Croatia }
\maketitle

\begin{abstract}
We outline the twist-two analysis of deeply virtual Compton
scattering (DVCS) within the conformal partial wave expansion of
the amplitude, represented as a Mellin--Barnes integral. The
complete next-to-leading order results, including evolution, are
obtained in the $\overline{\mbox{MS}}$ and a conformal
factorization scheme. Within the latter, exploiting conformal
symmetry, the radiative corrections are evaluated up to
next-to-next-to-leading order. Using a new proposed
parameterization for GPDs, we study the convergence of
perturbation theory and demonstrate for H1 and ZEUS measurements
that our formalism is suitable for a fitting procedure of DVCS
observables. We comment on a recent claim of a
breakdown of collinear factorization and show that a
Regge-inspired $Q^2$ scaling law is ruled out by  small $x_{\rm
Bj}$ DVCS data.
\end{abstract}

\section{Introduction}

The proton structure has been widely explored in inclusive
measurements, mainly in deeply inelastic lepton-proton scattering
(DIS). Here the scattering essentially occurs due to the exchange
of a virtual boson (photon) between lepton and a single parton,
and so one can access \emph{parton distribution functions} (PDFs).
These universal, however, convention-dependent functions
$q_{a}(x)$ are interpreted as probabilities that partons of
certain flavour $a$ will be found with given longitudinal momentum
fraction $x$. Since the PDFs are naturally defined in a
translation invariant manner, they do not carry information about
the transversal distribution of partons. Some information about
transversal degrees of freedom can be obtained from elastic
lepton-proton scattering. Namely, the electromagnetic form factors
$F_{1,2}(t)$ are Fourier transforms of the electric and magnetic
charge distribution in nucleon, and can be, e.g., in the infinite
momentum frame, interpreted as probability that partons are found
at some transversal distance ${\bf b}$ from the center-of-mass.
However, one should not assume that a realistic probability
distribution of partons, given by a two-variable function $q(x,
{\bf b})$ is simply a direct product, i.e., $q(x) \otimes q({\bf
b})$, of two probability functions. Rather it is anticipated that
longitudinal and transversal degrees of freedom have a cross talk,
e.g., as $x$ gets bigger partons carry more of the nucleon
longitudinal momentum and are expected to be closer to the proton
center, and thus the ${\bf b}$ dependence in $q(x,{\bf b})$ should
become narrower with increasing $x$.

The three dimensional distribution of partons in the nucleon can
be addressed within more general objects, called \emph{generalized
parton distributions} (GPDs) \cite{MueRobGeyDitHor94,Rad96,Ji96}.
Such distributions can be revealed by analyzing  hard exclusive
leptoproduction of mesons or photon. The theoretical description
of the former processes is perhaps more problematic, but they
offer a direct view into individual flavour GPDs. To the latter
one the \emph{deeply virtual Compton scattering} (DVCS) process
contributes, where one photon has a large virtuality. DVCS is
theoretically considered as the cleanest probe of GPDs, however,
here only certain flavour combinations of GPDs appear.

The non-forward Compton scattering process is described by the
Compton tensor
\begin{eqnarray}
\label{Def-ComScaTen} T_{\mu\nu} (q, P, \Delta) = \frac{i}{e^2}
\int\! d^4x\, e^{i x\cdot q} \langle P_2, S_2 | T j_\mu (x/2)
j_\nu (-x/2) | P_1, S_1 \rangle,
\end{eqnarray}
where $q = (q_1 + q_2)/2$, $P = P_1 + P_2$ and $\Delta=P_2-P_1$.
The generalized Bjorken limit corresponds to
$Q^2 = -q^2\to \infty$ with the scaling
variables
\begin{eqnarray}
\xi = \frac{Q^2}{P\cdot q}\,,\qquad \eta = -\frac{\Delta\cdot
q}{P\cdot q}\,,
\end{eqnarray}
and the momentum transfer squared $\Delta^2$  being fixed. Note
that in the forward case, i.e., $\Delta \to 0$, the hadronic DIS
tensor $W_{\mu\nu}$ is related to the \emph{forward} Compton
scattering tensor by the optical theorem
\begin{equation}
W_{\mu\nu} = \Im{\rm m}
T_{\mu\nu} (q,P=2 p,\Delta=0)/(2 \pi) \, ,
\end{equation}
where $p=P_1=P_2$ and $\xi \to x_{Bj}$.

\begin{figure}
\begin{tabular}{ccc}
\includegraphics[scale=0.7]{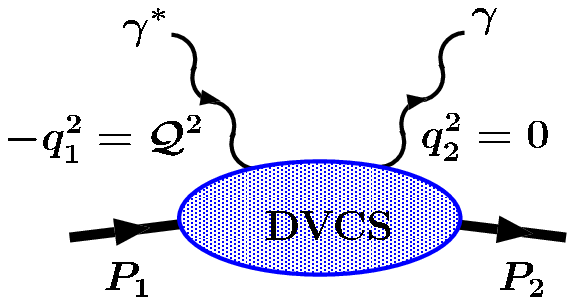}
&
\includegraphics[scale=0.7]{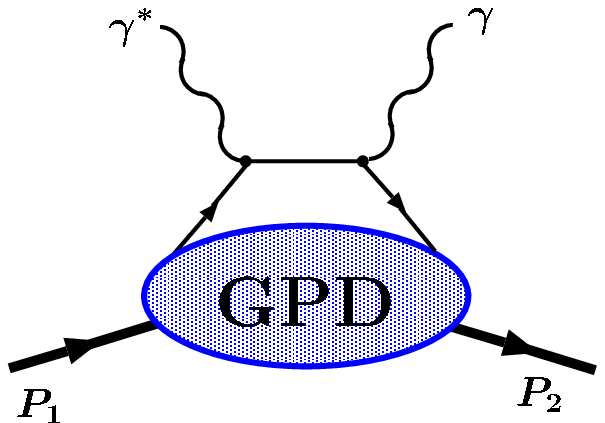}
&
\includegraphics[scale=0.7]{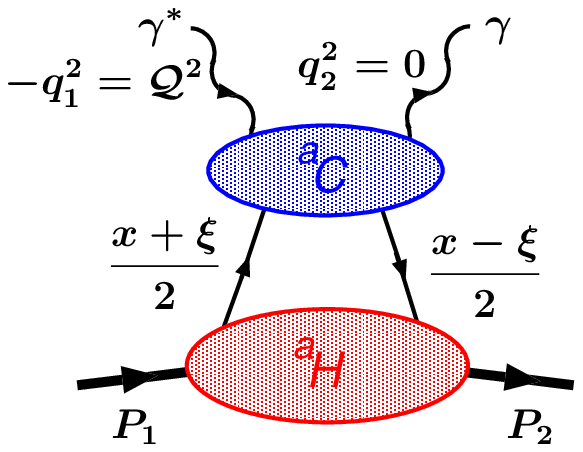}
\\
a)
&
b)
&
c)
\end{tabular}
\caption{a) DVCS.
         b) Leading-order perturbative contribution to DVCS.
         c) Factorization (to all orders in $\alpha_s$) on an example of
            special parity-even, helicity conserving contribution
            $^a{\cal H}$, Eq. (\protect\ref{eq:conv}).}
\label{fig:DVCS}
\end{figure}

In DVCS -- Fig.~\ref{fig:DVCS}a -- the virtuality of the incoming
photon ${\cal Q}^2=-q_1^2$ is large while the final photon is
on-shell. The skewness parameter $\eta$ and the Bjorken-like
scaling parameter $\xi$ are then equal to twist-two accuracy,
i.e., $\eta = \xi + {\cal O}(1/{\cal Q}^2)$. In the generalized
Bjorken limit, similarly as for DIS structure functions, the
amplitude factorizes \cite{Radyushkin97,*JiO98,*CollinsF98} into
long- and short-distance contributions (Fig.\ref{fig:DVCS}b and
c): short-distance physics is perturbatively calculable Compton
scattering on a parton, while long-distance physics is encoded in
a non-perturbative amplitude for a parton being emitted and later
reabsorbed by the nucleon. The latter {\em amplitude} is called
GPD. The factorization of long- and short-distance contributions
is based on the reshuffling of divergent contributions from the
perturbative calculable hard-scattering part to the universal,
i.e., process independent GPDs. Although GPDs are universal they
depend on the scheme convention and also on the order of the
approximation to which one is able to perform the perturbative
calculation.

The GPDs with even parity, considered  here, are defined as
\begin{eqnarray}
\label{Def-GPD-Q}
 {}^q\! F (x, \eta, \Delta^2)& = &
  \int \frac{d z^-}{2 \pi}\; e^{i x P^+ z^-}
  \langle {P_{2}}|{\bar{q}(-z)\gamma^+ q(z)}| {P_{1}} \rangle
  \Big|_{z^+=0,\, \mathbf{z}_\perp=\mathbf{0}}
\, , \\
\label{Def-GPD-G}
 {}^G\! F (x, \eta, \Delta^2)& = &
  \frac{4}{P^+} \int \frac{d z^-}{2 \pi}\; e^{i x P^+ z^-}
  \langle P_2|{G^{+\mu}_{a}(-z) G_{a \mu}^{\;\;\: +} (z)}|P_1\rangle
\Big|_{z^+=0,\, \mathbf{z}_\perp=\mathbf{0}}
\, ,
\end{eqnarray}
and similarly for odd parity. Furthermore, it is convenient to
make a decomposition,
\begin{equation}
  ^a\! F = \frac{\bar{u}(P_2) \gamma^+ u(P_1)}{P^+} \, {{}^a\!H} +
 \frac{\bar{u}(P_2) i \sigma^{+\nu} u(P_1) \Delta_\nu}{2 M P^+}\, {{}^a\!E}
\, ,
     \qquad a=q, G
\, ,
\end{equation}
into helicity conserving and non-conserving generalized form
factors. The Compton tensor (\ref{Def-ComScaTen}) is analogously
decomposed into Compton form factors with corresponding parity and
helicity properties. In the forward limit ($\Delta \to 0$) GPDs
reduce to PDFs. Together with sum rules, e.g., relating GPDs to
electromagnetic form factors, this provides constraints that are
important for GPD modelling. But as they do not constrain much the
skewness ($\eta$) dependence, modelling is still a difficult
problem. One guiding feature is a polynomiality property of GPDs:
$n$-th (Mellin) moment of GPD is even polynomial in $\eta$ of
order $n$ or $n\pm 1$. Similar polynomiality will be obeyed also
by conformal moments which are just linear combination of Mellin
moments. The usefulness of the GPDs has also been widely realized
in connection with the spin problem, since they encode the angular
momentum carried by the individual parton species, as explicated
by the Ji's sum rule \cite{Ji96}. For recent detailed account of
GPDs and their properties, we refer to
\cite{Diehl03,*BelitskyR05}.

In the following we first briefly comment on the work in
Ref.~\cite{SzcLonLla07}, which claims that
collinear factorization breaks down. We then proceed with the outline of the
perturbative QCD approach to DVCS that is based on the conformal
partial wave expansion, represented as Mellin-Barnes integral and
described in detail in Refs.~\cite{MuellerSch05,KumerickiMPK07}. Here, as in
Refs.~ \cite{Mueller05,KumerickiMPKSch06,KumerickiMPK07}, we
concentrate on the dominant Compton form factor ${\cal H}$
corresponding to parity-even helicity-conserving GPD $H$. For
simplicity, we present only the results for the singlet part which
is dominant for the kinematics of collider experiments. We shortly
discuss our phenomenological analysis of H1 and ZEUS DVCS data.
Finally, we conclude.

\section{Does collinear factorization break down?}

Here we would like to comment on a recent claim of an observed breakdown
of collinear factorization \cite{SzcLonLla07}, which the authors
consider as a general feature of deeply exclusive leptoproduction.
This claim originated from a divergent result that was found by convoluting a
model GPD with the hard-scattering part in the handbag approximation,
cf.~Fig.~\ref{fig:DVCS}b.
In our opinion the GPD model employed by Ref. \cite{SzcLonLla07} suffers
from ultraviolet and/or infrared
divergencies. A matching procedure  with the collinear
factorization approach is impossible, since the model does not respect
the ultraviolet behavior of QCD.

\begin{itemize}
\item
Divergences in a non-perturbative model that can not be
cured within the factorization approach have to be removed by
hand, e.g., by an effective cut off, before the model can be utilized.
Convoluting a model GPD, even a finite one, with the hard-scattering
part, and obtaining undefined result does not in itself constitute
a proof of a breakdown of collinear factorization.
\end{itemize}

\noindent We recall that the collinear factorization approach has
been worked out for deeply exclusive leptoproduction to NLO
\cite{Ji97,*BelitskyM97,*MankiewiczPSVW97,IvaSzyKra04}, including
evolution, and its quantitative features have been extensively studied. Below
we report about our own investigations of DVCS beyond this order.

We find it interesting that the analysis of
Ref.~\cite{SzcLonLla07} resulted in an infinite DVCS amplitude.
Since it was claimed that this is a necessary consequence of Regge
behavior, we would like to offer our opinion about the origin of
these divergences. Let us remind that the field-theoretical
definition (\ref{Def-GPD-Q}) of GPDs states that they are
generalized functions in the mathematical sense rather then
regular ones \cite{MueRobGeyDitHor94}. To shorten the discussion
and without loos of generality, we will employ this fact for the
evaluation of the convolution integral \cite{GelShi64}.

The non-perturbative dynamics of the GPD utilized in
Ref.~\cite{SzcLonLla07} is described by a parton model. It is
supposed that the collective spectator  has  leading Regge
behavior, i.e., $0<\alpha<1$, and interacts {\em point-likely}
with the active {\em spin-zero} constituents. This reduction of the
spin content induces a factor $x$ which is taken outside of the
GPD and leads to a  {\em non-standard} representation
\cite{BelKirMueSch01}. This causes a mathematical paradox for
``good'' GPDs. Namely,  the inverse, i.e., $1/x$, moment of a GPD
is defined, contrarily to what one expects on general grounds,
while in the forward limit the inverse  moment of the resulting
PDF is divergent for $0<\alpha<1$. Obviously, to be consistent,
a {\em non-standard} GPD must contain some ``bad'' terms.
The solutions are (i.)~to reject such a GPD
model, (ii.)~to put the ``bad'' part in the central region $-\eta \le x \le \eta$, where it serves
as a counterterm, and (iii.)~to define the inverse
moment for the PDF.
Our choice in Eq. (\ref{Def-H}) below corresponds to the last possibility
and the illustration of thus obtained ``good'' GPDs.
Furthermore, we stress that the {\em point-like} coupling of the
constituents  ties the high-energy and the ultraviolet behavior.
The treatment of divergencies\footnote{In the model one can choose
to have ultraviolet, infrared, or both kinds of
divergencies, however, one can not get rid of all of them. After
momentum integration the GPDs will always suffer from the same
illness, which is  not primarily related to the high-energy
behavior, rather, it is the generic feature of the underlying
simplifications.} as part of modelling was not specified
in Ref.~\cite{SzcLonLla07} and,
consequently, the resulting GPDs contain ambiguous terms.

\begin{itemize}
\item
We consider the GPD model  \cite{SzcLonLla07} in its presented form
as mathematically ill-defined.
\end{itemize}

An elegant method to overcome all these obstacles is to consider
the GPD model \cite{SzcLonLla07} as an analytic function of the
parameter $\alpha$. This allows removal of ultraviolet (or
infrared) divergencies, except at discrete values of $\alpha$,
e.g., $\alpha=0$ and $\alpha=1$. Possible version of  the GPD model
part that originates from the leading Regge behavior reads
in the region $-\eta \le x \le 1$ as follows
\begin{eqnarray}
\label{Def-H}
H(x,\eta) = - \frac{N}{2\alpha}\frac{x}{\eta} \left[ \theta(x+\eta) \left(\frac{x+\eta}{1+\eta}\right)^{-\alpha} -
\theta(x-\eta) \left(\frac{x-\eta}{1-\eta}\right)^{-\alpha}\right]\,.
\end{eqnarray}
Up to terms which live only in the central region $-\eta \le x \le
\eta$, and Regge non-leading terms, this expression coincides with
the result given in Ref.~\cite{SzcLonLla07}. Note that we consider
$(\cdots)^{-\alpha}$ as generalized functions, which might be
expressed using conventional ``+'' definitions and some additional
subtraction terms. The forward limit yields the PDF $H(x,\eta=0) =
N x^{-\alpha}(1-x)$ and its inverse moment, i.e.,
$\int_{(0)}^1\!dx\, x^{-1-\alpha}(1-x)=1/\alpha (\alpha-1)$,  is
defined. Armed with a clear mathematical language \cite{GelShi64}, we calculate
the convolution with the hard scattering part and get a finite
result, however, with an `unusual' negative overall sign:
\begin{equation}
{\cal H}(\xi) = \int_{-1}^1\!dx\, \frac{2x}{\xi^2-x^2-i \epsilon} H(x,\xi) =
-\frac{N \pi}{2\alpha} (2\xi)^{-\alpha} \left[i - \cot\left({\frac{\pi \alpha}{2}}\right) +O (\xi) \right] + O(\xi^0)
\,.
\end{equation}

\begin{itemize}
\item
The reported divergences in Ref.~\cite{SzcLonLla07} do {\em not}
show up in the leading Regge behavior. They arise
from ill-defined subtractions, entirely related to
non-leading Regge behavior.
\end{itemize}

\noindent
Our inspection revealed that the ambiguous part in the GPD model \cite{SzcLonLla07},
affected by ultraviolet and/or infrared divergences, possesses non-leading Regge
behavior. It shows up only in the central GPD region  and causes the
observed divergences in the DVCS amplitude.

The fact that the GPD representation is not unique was
pointed out in Refs.~\cite{Ter01,BelKirMueSch01}. We recall that
numerous QCD model evaluations confirm within the {\em standard}
GPD definition a finite and continuous behavior at the cross-over
points $x=\pm\eta$, see
Refs.~\cite{Rad97,*Rad00a,*GoePolVan01,Diehl03,*BelitskyR05} and
references therein.
In contrast to the representation used in
Ref.~\cite{SzcLonLla07}, Regge behavior of standard
GPDs is continuous at the cross-over points, e.g., for
$\alpha < 1$:
\begin{eqnarray}
\label{Tra-for} \frac{x}{\eta}\left[
\left(\frac{x-\eta}{1-\eta}\right)^{-\alpha}\!\!\! -\left(\frac{x+\eta}{1+\eta}\right)^{-\alpha}
\right]\,,
&&
\frac{1}{\eta}\left[\left(\frac{x+\eta}{1+\eta}\right)^{1-\alpha} \!\!\!-
\left(\frac{x-\eta}{1-\eta}\right)^{1-\alpha}\right] \;
\mbox{for}\;  x \to \eta\,,\;\; \eta \le x  \,.
\nonumber\\
{} &&\nonumber
\\
\mbox{spin-0 partons}\qquad\quad\quad
&&
\qquad\qquad
\mbox{spin-1/2 partons}
\end{eqnarray}
If one chooses the standard GPD representation, the $s$-channel view
in Ref.~\cite{SzcLonLla07} confirms the implementation of Regge
behavior in  common GPD models \cite{Rad97,*Rad00a,*GoePolVan01}
and our approach.  To see this, one simply convolutes the spectral
function (9) of Ref.~\cite{SzcLonLla07} and the models (19) and
(21) of Ref.~\cite{HwaMue07} with respect to the spectator mass
$\lambda$. Doing so one finds the  r.h.s.~in (\ref{Tra-for}).

Finally, in Ref.~\cite{SzcLonLla07} it was suggested  to
utilize so-called Regge-exchange amplitudes for phenomenology.
These amplitudes are not universal and possess a typical $s^{\alpha(t)}\sim ({\cal
Q}^2/\xi)^{\alpha(t)}$ Regge behavior, which yields a softening of
Bjorken scaling. Let us recall that Regge phenomenology
has been developed out of $S$-matrix theory, i.e., for processes with
asymptotic states, e.g., $q_1^2=q_2^2=0$, in lack of a dynamical
understanding of strong interaction phenomena. Hence, one might
surmise that the implementation of Regge-behavior in a parton
model within virtual constituents and {\em point-like} interaction
may predict unrealistic scaling for off-shell amplitudes. A last
objection arises when the claimed scaling behavior
\cite{SzcLonLla07} is confronted with DVCS data in the high-energy region
(see Sect. 5 below):

\begin{itemize}
\item
The  generic $({\cal Q}^2/\xi)^{\alpha(t)}$ behavior of Regge-exchange
amplitudes  is ruled out by  experiment.
\end{itemize}

\noindent We emphasize that this statement is based on
experimental data, presently available. Our investigation, given
below, will show where perturbation theory reveals itself and in
which kinematics the high-energy limit spoils the Bjorken limit.
It turns out that this is rather  a universal feature, which is
tied to the ultraviolet behavior in QCD.


\section{Deeply virtual Compton scattering}

Besides DVCS, the Bethe-Heitler (BH) brehmstrahlung process
contributes to the measured hard photon leptoproduction off a
proton. The BH amplitude, known in leading order of the  QED fine
structure constant, is expressed in terms of the known
electromagnetic form-factors. Generally, there are two types of
DVCS experiments: collider experiments and fixed target
experiments. The former usually provide information in the phase
space $10^{-4} \lesssim \xi \lesssim 10^{-1}$ and $1\,
\mbox{GeV}^2 \lesssim {\cal Q}^2 \lesssim 100\, \mbox{GeV}^2$, and
such are H1 and ZEUS experiments at HERA. The main observables
here are total and differential DVCS cross sections, however, also
the measurement of beam charge asymmetry is feasible. For the
fixed target experiments, such as Hall A, Hall B (CLAS) at JLAB,
and HERMES at DESY, the DVCS--BH interference term can be more
easily accessed via single beam, target spin  and beam charge
(HERMES) asymmetries, while the investigated phase space covers
the so-called valence quark region, i.e., $0.05 \lesssim \xi
\lesssim 0.3$ within $1\, \mbox{GeV}^2 \lesssim {\cal Q}^2\lesssim
10\, \mbox{GeV}^2$.

One can express ${\cal H}$ as a convolution (Fig.\ref{fig:DVCS}c)
over the longitudinal momentum fraction $x$
\begin{equation}
{}^{a}\mathcal{H}(\xi, \Delta^2, \mathcal{Q}^2) =
\int {\rm d}x\; ^{a}\!C (x, \xi, \mathcal{Q}^2/\mu^2) \;
{^{a}\! H (x, \eta=\xi, \Delta^2, \mu^2)}
\, ,
\label{eq:conv}
\end{equation}
where $\mu^2$ is a factorization scale that separates short- and
long-distance dynamics and is often taken as
$\mu^2=\mathcal{Q}^2$. Here the index $a\in\{\mbox{NS,S}(\Sigma,
G)\}$ denotes either non-singlet or singlet parts, where to latter
both quarks ($\Sigma$) and gluons ($G$) contribute.

The coefficient functions $C^a$ are perturbative quantities which
describe $q \gamma^* \to q \gamma$ and $g \gamma^* \to g \gamma$
subprocesses. The well known leading-order (LO) contribution to
$C^a$ is actually a pure QED process (Fig. \ref{fig:DVCS}b). The
next-to-leading order (NLO) contribution --- the first order in
$\alpha_s$ --- has been calculated by various groups
\cite{Ji97,*BelitskyM97,*MankiewiczPSVW97}. Obviously, to
stabilize the perturbation series and investigate its convergence
one needs the second order in $\alpha_s$, i.e.,
next-to-next-to-leading order (NNLO) contributions. The importance
of NNLO in singlet case is amplified by the fact that at LO
photons scatter only off charged partons, whereas gluons start
contributing  at NLO.

The GPDs ${}^{a}\!H$ are intrinsically non-perturbative quantities
whose form at some initial scale $\mathcal{Q}_0$ has to be deduced
by some non-perturbative methods (lattice calculation, fit to
data, etc.). The evolution to the factorization scale of interest
is governed by perturbation theory,
cf.~\cite{KumerickiMPKSch06,KumerickiMPK07} for LO examples. The
anomalous dimensions of non-diagonal operators were calculated up
to NLO \cite{Mueller93,*BelitskyM98,*BelitskyM99}. Still,
evolution at NLO is  numerically not easy to implement, and has
been investigated beyond NLO only recently, using the procedure
explained below \cite{KumerickiMPK07}.

Instead of using the convolution (\ref{eq:conv}) one can
equivalently use the sum over the conformal moments, and the
singlet contribution then takes the form
\begin{equation}
   {^{\rm S}\! {\cal H}}(\xi,\Delta^2,{\cal Q}^2)
= 2 \sum_{j=0}^\infty \xi^{-j-1}
\mbox{\boldmath $C$}_{j}({\cal Q}^2/\mu^2,\alpha_s(\mu))\;
\mbox{\boldmath $H$}_{j}(\eta=\xi,\Delta^2,\mu^2)
\, ,
\label{eq:sumcm}
\end{equation}
where $\mbox{\boldmath $C$}_j=\left( ^\Sigma\! C_j, ^G\!\! C_j
\right)$ and $\mbox{\boldmath $H$}_j=\left( ^\Sigma\!H_j, ^G\!\!
H_j \right)$ are conformal moments. They are analogous to common
Mellin moments used in DIS but the integral kernel $x^j$ is
replaced by Gegenbauer polynomials $C_j^{3/2}(x)$ and
$C_j^{5/2}(x)$, which are solutions of LO evolution equations for
quarks and gluons, respectively. Unfortunately, the series
(\ref{eq:sumcm}) only converges in the unphysical region. Hence,
it is necessary to resum this series, e.g., by means of the
Mellin-Barnes integral \cite{KumerickiMPK07}
 \begin{equation}
 {^{\rm S}\!{\cal H}}(\xi,\Delta^2,{\cal Q}^2)
 = \frac{1}{2i}\int_{c-i \infty}^{c+ i \infty}\!
 dj\,\xi^{-j-1} \left[i +
 \tan
 \left(\frac{\pi j}{2}\right) \right]
 \mbox{\boldmath
 $C$}_{j}({\cal Q}^2/\mu^2,\alpha_s(\mu))
 \mbox{\boldmath $H$}_{j}(\xi,\Delta^2,\mu^2)
\, .
\end{equation}

The advantages of using conformal moments, i.e., Mellin-Barnes
representation, are manifold: it enables a simple inclusion of
evolution, it allows for an efficient and stable numerical
treatment, and it opens a new approach to modelling of GPDs.
Finally, by making use of conformal operator product expansion
(OPE) and known NNLO DIS results, it enables the assessment of
NNLO contributions, to which we now turn.

\section{Conformal approach to DVCS beyond NLO}

Neither Wilson coefficients nor anomalous dimensions are
calculated in non-forward kinematics at NNLO (only so-called quark
bubble insertions were partly evaluated
\cite{BelSch98,*MelNicPas02}). To access the NNLO of non-forward
Compton scattering, we use the conformal approach, making it
possible to calculate relevant objects using only diagonal results
of forward Compton scattering, i.e., DIS.

DVCS belongs to a class of two-photon processes (DIS, DVCS,
two-photon production of hadronic states \ldots) calculable by
means of the OPE, $T_{\mu\nu} (q, P, \Delta) \rightarrow C_j \,
O_j$, for which the use of generalized Bjorken kinematics and
conformal symmetry enables a unified description. While massless
QCD is conformally invariant at tree level, this invariance is
broken at the loop level since renormalization introduces a mass
scale, leading to the running of the coupling constant ($\beta \ne
0$). Assuming the existence of a non-trivial fixed point
$\alpha_s^\ast$, i.e., $\beta(\alpha_s^\ast)=0$, the conformal OPE
(COPE) prediction for Wilson coefficients in general kinematics
reads \cite{Mueller97,*Mueller98,*MelicMPK02}
\begin{eqnarray}
C_j(\alpha_s^\ast)
&=&  c_j(\alpha_s^\ast)
\, {_2F_1}\!\!\left(\!\!{(2+2 j
+\gamma_j(\alpha_s^\ast))/4, (4+2 j + \gamma_j(\alpha_s^\ast))/4
\atop (5+2 j +
\gamma_j(\alpha_s^\ast))/2}\Big|\frac{\eta^2}{\xi^2}\! \right)
\left(\!\!\frac{\mu^2}{Q^2}\!\!\right)^{\gamma_j(\alpha_s^\ast)/2}\,.
\qquad
\label{eq:cope}
\end{eqnarray}
For $\eta=0$ equation (\ref{eq:cope}) reduces to the DIS Wilson
coefficients $C_j \to c_j$ and thus fixes the normalization $c_j$.
The choice $\eta=\xi$ corresponds to DVCS in the conformal limit.
The anomalous dimensions governing the evolution are diagonal and
the same as in DIS.

For a general factorization scheme, e.g., the $\overline{\rm MS}$
scheme, the conformal symmetry breaking occurs also due to the
renormalization of the composite operators and causes the
appearance of non-diagonal anomalous dimensions $\gamma_{jk} =
\delta_{jk} \gamma_j + {\gamma_{jk}^{\rm ND}}$ . This induces a
mixing of both operators, i.e., GPDs, and Wilson coefficients
under evolution. In the kinematical forward limit ($\eta=0$) the
diagonal evolution equations are again obtained, i.e., the DIS
case corresponds to the COPE result. For DVCS, evaluated in the
$\overline{\mbox{MS}}$ scheme, there appear also conformal
symmetry breaking terms which are not proportional to $\beta$,
i.e., the  non-diagonal terms survive.

The non-diagonal terms of anomalous dimensions encountered in
$\overline{\mbox{MS}}$ scheme can be removed by a finite
renormalization \cite{Mueller97,*Mueller98,*MelicMPK02}, i.e, by a
specific choice of the factorization scheme
\begin{equation}
C^{\overline{{\rm MS}}}\, O^{\overline{{\rm MS}}} = C^{\overline{{\rm MS}}} \, B \,
B^{-1} \, O^{\overline{{\rm MS}}} = C^{\overline{{\rm CS}}} O^{\overline{{\rm CS}}}.
\end{equation}
In this new scheme, called conformal subtraction
($\overline{\mbox{CS}}$) scheme, all non-diagonal terms are
''pushed'' to the $\beta$ proportional part $
\gamma^{\overline{{\rm CS}}}_{jk} = \delta_{jk} \gamma_k  { +
\beta/g \Delta_{jk} } $. Furthermore, since there is an ambiguity
in $\overline{\mbox{MS}} \to \overline{\mbox{CS}}$ rotation
matrix, by judicious choice {$\delta B$} one can ``push'' mixing
under evolution to NNLO. Hence, in $\overline{\mbox{CS}}$ scheme,
which we are using, the unknown correction $\Delta_{jk}$ starts at
NNLO and it can be additionally suppressed by the choice of an
appropriate initial condition. Finally, we express our result in
$\overline{\mbox{CS}}$ scheme as
\begin{eqnarray}
C_j^{\overline{{\rm CS}}}
  &=& \sum_{k=j}^{\infty}{C_k(\alpha_s({\cal Q}))} \;
{\cal P}
\exp\left\{\int^\mu_{\cal Q}\frac{d\mu^\prime}{\mu^\prime}
\left[\gamma_j(\alpha_s(\mu^\prime))
\delta_{kj}+ \frac{\beta}{g}
\Delta_{kj}(\alpha_s(\mu^\prime))\right]\right\}\,,
\end{eqnarray}
with $C_k(\alpha_s({\cal Q}))$ obtained from the $\eta=\xi$ limit
of Eq. (\ref{eq:cope})  and using $c_{j}(\alpha_s)$. As stated
above the $\Delta_{kj}$ mixing term appears at NNLO and is
neglected. We take $c_{j}$ and  $\gamma_j$ calculated to NNLO
order from Refs.~\cite{ZijlstravN92,*VogtMV04}, and obtain the
DVCS prediction to NNLO in the $\overline{\mbox{CS}}$ scheme.

\section{Results}
\begin{figure}
\centerline{\includegraphics{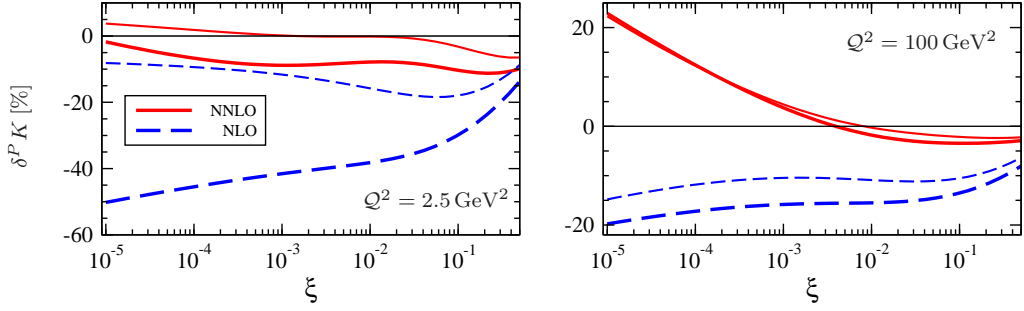}}  %
\caption{
Relative NLO and NNLO corrections
(\protect\ref{eq:relcorr})
 in the $\overline{\mbox{CS}}$ scheme
($\Delta^2=0.25$GeV$^2$, input scale ${\cal Q}_0^2=2.5$GeV$^2$).
Thick [thin] lines denote ``hard'' [``soft''] gluon scenarios:
$N_G = 0.4$,  $\alpha_{G}(0) = \alpha_{\Sigma}(0)+ 0.05$
[$N_G = 0.3$,  $\alpha_{G}(0)= \alpha_{\Sigma}(0)-0.02$].
}
\label{f:NNLO}
\end{figure}
We have used the formalism described in the preceding sections to
investigate the size of NNLO corrections to non-singlet
\cite{Mueller05} and singlet Compton form factors
\cite{KumerickiMPKSch06}, to obtain complete (non-diagonal
evolution included) $\overline{\mbox{MS}}$ NLO predictions
\cite{KumerickiMPK07}, and to perform fits, in both
$\overline{\mbox{MS}}$ and $\overline{\mbox{CS}}$ schemes, to DVCS
and DIS data and to extract information about GPDs
\cite{KumerickiMPK07}.

One can use a simple Regge-inspired ansatz for GPDs at the input scale
\begin{equation}
\label{ansatz}
\mbox{\boldmath $H$}_j(\eta, \Delta^2, {\cal Q}_{0}^2) =
\left(
\begin{array}{c}
N'_{\Sigma}\;\! F_{\Sigma}(\Delta^2)\,
{{B}}\bigl(1+j-\alpha_{\Sigma}(0),8\bigr) \\
N'_{{G}}\;\! F_{{G}}(\Delta^2)\,
{{B}}\bigl(1+j-\alpha_{{G}}(0),6\bigr)
\end{array}
\right)
+ {\cal O}(\eta^2)
\, ,
\end{equation}
with ($p_a$ is a flavour dependent integer)
\begin{equation}
{\alpha_{a}(\Delta^2) = \alpha_{a}(0) + 0.15 \Delta^2} \, , \qquad
F_{a} (\Delta^2) = \frac{j+1-\alpha_a(0)}{j+1-\alpha_a(\Delta^2)}
\left(1 - \frac{\Delta^2}{ {M_{0}^{a}}^2 +j  {M_{\Delta}^{a}}^2 }\right)^{-p_a} \vspace*{2.5ex}
\, .
\end{equation}
Here $\alpha_a$ are effective Regge-trajectories and $F_a$ at
$j=0$ are partonic form factors, which are modelled as product of
a monopole factor, arising from the effective Regge exchange, and
an impact form factor. The $\Delta^2$-dependence of higher moments
could be easily pinned down from realistic lattice measurements,
see Ref.~\cite{Hagetal07} and references therein for the present
status of simulations in the flavour sector. The ansatz is based
on modelling the $t$-channel contributions of the GPDs in terms of
SO(3) partial waves \cite{PolShu02} at a low input scale and then
using analyticity, implicitly tied to Lorentz symmetry
(polynomiality of GPD moments), to extend the GPD support
\cite{MuellerSch05}. Our assumption is that at a rather low input
scale, i.e., photon virtuality, Regge behavior is present in the
GPD moments and can be effectively modelled by poles in the
complex conformal spin plane. Doing that in such a way yields GPDs
that behave continuously at the cross-over point and can so be
smoothly incorporated in the factorization approach. We emphasize
that this is compatible with the $s$-channel view
\cite{SzcLonLla07}, if it would be presented in an appropriate
representation, cf.~Eq.~(\ref{Tra-for}). In the forward case
($\Delta = 0$) ansatz (\ref{ansatz}) is equivalent to the standard
building blocks for PDFs:
\begin{equation}
\Sigma(x) = N'_{\Sigma}\,
x^{-\alpha_\Sigma(0)}\, (1-x)^7,\qquad G(x) = N'_{{G}} \,
x^{-\alpha_{{G}}(0)} \, (1-x)^5.
\end{equation}
We relied in our ansatz (\ref{ansatz}) only on the leading SO(3)
partial wave, which can be for  $\eta \lesssim 0.3$ safely
approximated by a constant. The work on a more flexible
$\eta$-dependent ansatz, i.e., the model dependent resummation of
SO(3) partial waves is in progress.

We have performed the analysis of radiative corrections with
generic parameters and made fits of relevant parameters
$N_{\Sigma}$, $\alpha_{\Sigma}(0)$, $M_{0}^\Sigma$, $N_G$,
$\alpha_{G}(0)$, $M_{0}^{G}$ (since of $j\sim 0$ dominance
$M^{a}_\Delta$ can be safely set to zero). We introduce now the
quantities that we utilize as a measures of the scheme dependence
and, foremostly, as indicators for the convergence of the
perturbation series. It is natural to employ for this purpose the
ratios of Compton form factors, i.e., the corresponding modulus
and phase difference, at order ${\rm N}^{P}{\rm LO}$ to those at
order ${\rm N}^{P-1}{\rm LO}$, where $P=\{0,1,2\}$ stands for LO,
NLO, and NNLO order, respectively:
\begin{equation}
\delta^{P} K=  \frac{\left|{{\cal H}}^{{\rm N}^P{\rm LO}}\right|}
{\left|{{\cal H}}^{{\rm N}^{P-1}{\rm LO}}\right|}
 -1\;,
 \qquad
\delta^P \varphi=
{\rm arg}\!\left( \frac{{\cal H}^{{\rm N}^P{\rm LO}}}{ {\cal H}^{{\rm N}^{P-1}{\rm LO}}}\right)
\, .
\label{eq:relcorr}
\end{equation}
The phase differences are small, and we will not comment on them
here further. The NLO corrections to the moduli in
$\overline{\mbox{MS}}$ and $\overline{\mbox{CS}}$ schemes have a
similar $\xi$-shape, where $\overline{\mbox{MS}}$ corrections are
generally larger. The relative NLO and NNLO corrections in
$\overline{\mbox{CS}}$ scheme are depicted in Fig.~\ref{f:NNLO}.
{From} the left panel, showing corrections at the input scale, we
realize that the large negative NLO corrections to the modulus in
the `hard' gluon scenario (thick dashed) are shrunk at NNLO to
less than $10\%$ (thick solid), in particular in the small $\xi$
region. In the `soft' gluon case the NNLO corrections (thin solid)
are  $\pm 5\%$. For $\xi\sim 0.5$, the corrections are reduced
only unessentially and are around $5\%$ and $10\%$ at both NLO and
NNLO level. If evolution is switched on (right panel), our
findings drastically change. For $  5\cdot 10^{-2} \lesssim \xi$ NNLO
corrections are stabilized on the level of about $3\%$ at ${\cal
Q}^2 = 100\ \mbox{GeV}^2$. But they start to grow with decreasing
$\xi$ and reach at $\xi\approx 10^{-5}$ the $20\%$ level. As in
DIS, this breakdown of perturbation theory at small $\xi$ in DVCS
obviously stems from evolution and is thus universal, i.e.,
process independent. The large change of the scaling prediction
within the considered order does not influence the quality of
fits, and, in particular, the possibility of relating DVCS and DIS
data. Hence, the problem of treatment or resummation of these
large corrections is relevant primarily to our partonic
interpretation of the nucleon content. As long as we precisely
define the treatment of the evolution operator, perturbative QCD
can be employed as a tool for analyzing data also in the small
$\xi$ region. However, it might be expected that in meson
leptoproduction dominant perturbative corrections
\cite{IvaSzyKra04,DieKug07} at small $\xi$ in the hard scattering
amplitude are not connected only to the evolution and should be
resummed \cite{IvaPri07}.

\begin{figure}[t]
\centerline{\includegraphics[scale=0.8]{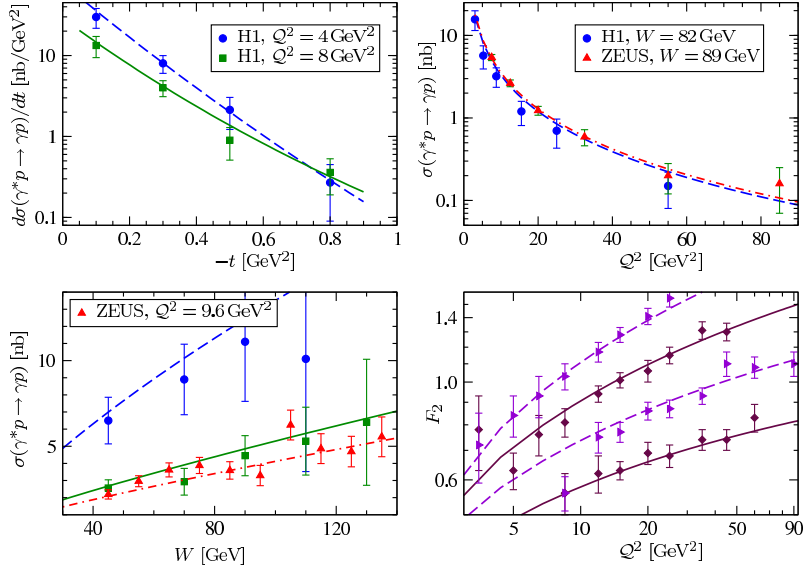}}
\vspace{-0.2cm} \caption{Simultaneous fit to the DVCS and DIS data
in the $\overline{\rm CS}$  scheme to NNLO. Upper left panel DVCS
cross section  for ${\cal Q}^2 =4\,{\rm GeV}^2$ and $W=71\, {\rm
GeV}$ (circles, dashed) as well as ${\cal Q}^2 =8\,{\rm GeV}^2$
and $W=82\, {\rm GeV}$ (squares, solid)
[27]
.
Upper right panel DVCS cross section ($|\Delta^2| < 1\,{\rm
GeV}^2$) versus ${\cal Q}^2$ for $W=82\, {\rm GeV}$ (H1, circles,
dashed) and $W=89\, {\rm GeV}$ (ZEUS, triangles, dash-dotted)
[28].
Lower left panel DVCS cross section versus
$W$ ${\cal Q}^2 = 4\,{\rm GeV}^2$ (H1, circles, dashed), ${\cal
Q}^2 = 8\, {\rm GeV}^2$ (H1, squares, solid), and ${\cal Q}^2 =
9.6\,{\rm GeV}^2$ (ZEUS, triangles, dash-dotted). Lower right
panel shows $F_2(x_{\rm Bj},{\cal Q}^2)$ versus ${\cal Q}^2$ for
$x_{\rm Bj}=\{8\cdot 10^{-3}, 3.2\cdot 10^{-3}, 1.3\cdot 10^{-3},
5\cdot 10^{-4}\}$
[29].}
\label{f:examplefit}
\end{figure}

The  Mellin-Barnes integral approach offers the possibility for a
fast and numerically stable analysis. Our numerical routine is
designed for the purpose of fitting DVCS (and DIS) observables and
testing various GPD ansaetze.  A fit example for measurements of
the H1 and ZEUS collaborations
\cite{Adletal01,Aktas:2005ty,Chekanov:2003ya,Aidetal96} is
presented in Fig.~\ref{f:examplefit} at NNLO. In these collider
experiments the DVCS cross section can be accessed and is given,
up to $\xi$ and $\langle  -\Delta^2  \rangle\!/4 M^2 \sim 0.05$
proportional corrections, at leading twist as
\begin{eqnarray}
\label{Def-CroSec1}
\frac{d\sigma}{d\Delta^2}(W,\Delta^2,{\cal Q}^2) \approx
\frac{4   \pi \alpha^2 }{{\cal Q}^4}\xi^2
\left| {\cal H} \right|^2
\left(\xi=\frac{{\cal Q}^2}{2 W^2+{\cal Q}^2},\Delta^2,{\cal Q}^2\right)\,.
\end{eqnarray}
For fixed $\xi$ we have the canonically expected scaling behavior
$1/{\cal Q}^4$, which is strongly modified by the resummation of
$\log{\cal Q}^2$ terms.

As one can realize from Fig.~\ref{f:examplefit} the quality of a
simultaneous DVCS and DIS fit within our ansatz (\ref{ansatz}) is
quite good, namely, $\chi^2/{\rm d.o.f.}=0.77$. We only observed
bad fits to LO accuracy, which originate from the fact that the
normalization of the DVCS amplitude  for fixed $\Delta^2$ is not
adjustable, since we took only the leading SO(3) partial wave. We
already demonstrated that taking the next-to-leading SO(3) partial
wave leads to good LO fits \cite{JLAB}. Note that this is nothing
else but a minimal version of the so-called dual GPD
parameterization \cite{PolShu02}, used in Ref.~\cite{GuzTec06}.
Let us emphasize that our fits support the following functional
form of CFFs in the small $\xi$-region:
\begin{eqnarray}
\left|  {\cal H}(\xi,\Delta^2,{\cal Q}^2)\right| \sim
N(\xi, {\cal Q}^2,\Delta^2)\,
\left(\frac{\xi_0}{\xi}\right)^{\alpha^{\rm eff}(\Delta^2,Q^2)}\,,
\end{eqnarray}
where $N(\xi, Q^2,t)$ and $\alpha^{\rm eff}(t,Q^2)$ effectively
describe the change of the normalization and slope due to the
predicted scaling violations, i.e., mainly arising from evolution.

In contrast, the so-called Regge-exchange amplitude \cite{SzcLonLla07}
yields a behavior which is
rather similar to the Regge behavior of on-shell amplitudes ($W^2\equiv s$)
\begin{eqnarray}
 \left| {\cal H}(\xi,\Delta^2,{\cal Q}^2)\right| \sim  N^\prime(\xi,\Delta^2)\,
 \left(\frac{W^2}{W^2_0}\right)^{\alpha(\Delta^2)}\,.
 \label{wrong-model}
\end{eqnarray}
Here the normalization factor $N^\prime(\xi,\Delta^2)$ can for
internal consistency of the model \cite{SzcLonLla07} only weekly
depend on $\xi \sim {\cal Q}^2/2 W^2$ and so it can be safely
neglected in the small $\xi$ kinematics. This prediction has so
far been compared only with exclusive leptoproduction data
\cite{Mor05,Cametal06,Airetal07} in the valence quark region
\cite{SzcLonLla07}, which is not conclusive \footnote{Note that
from present  fixed target data it is rather difficult to judge on
any predicted ${\cal Q}^2$ scaling behavior, since kinematical
variables are strongly correlated for a large ${\cal Q}^2$ lever
arm and, moreover, to access the twist-two sector in exclusive
meson leptoproduction  one has to extract the cross section that
arises from the exchange of a longitudinal polarized photon.
Within small ${\cal Q}^2$ lever arm and small $t$, DVCS
\protect{\cite{Cametal06}} indicates a canonical scaling which
might be logarithmical modified \protect{\cite{KumerickiMPK07}},
and is also still consistent with Regge scaling within a rather
low $\alpha(\ll\! t \!\gg)=0.15$  parameter
\protect{\cite{Diehl03,*BelitskyR05}}; as in DIS, exact canonical
scaling behavior perhaps takes place only in a very limited
$\xi$-region \protect{\cite{KumerickiMPK07}}.}. As spelled out
above, the DVCS amplitude can be described to LO accuracy, where
gluons are absent in the hard scattering part. Hence, for the sake
of illustration we consider it legitimate to confront the
Regge-exchange amplitude \cite{SzcLonLla07} within a pomeron Regge
trajectory $\alpha(\Delta^2) =1 + 0.25 \Delta^2 <1$
with high-energy DVCS data. As one can immediately realize
from Eqs.~(\ref{Def-CroSec1}) and (\ref{wrong-model}) for fixed
$W$ the cross section should not scale with ${\cal Q}^2$, which is
in conflict with the observation, describable by a $\left({\cal
Q}^2\right)^{-1.54\mp0.09\mp0.04}$ fit \cite{Aktas:2005ty}, as
shown in the upper right panel in Fig.~(\ref{f:examplefit}).
Confronting Eq.\ (\ref{wrong-model}) with the few experimental
data points for fixed $\xi$ \cite{Aktas:2005ty} also disfavors the
claimed Regge scaling. We note that a BFKL evaluation of the DVCS
amplitude indicates a rather intricate $\xi$ and ${\cal
Q}^2/\Delta^2$ dependencies in the high-energy limit
\cite{BalKuc00}, which seems not to support the conjectured Regge
behavior \cite{SzcLonLla07}, borrowed from on-shell amplitudes.

\begin{figure}
\begin{tabular}{cc}
\includegraphics[scale=0.62]{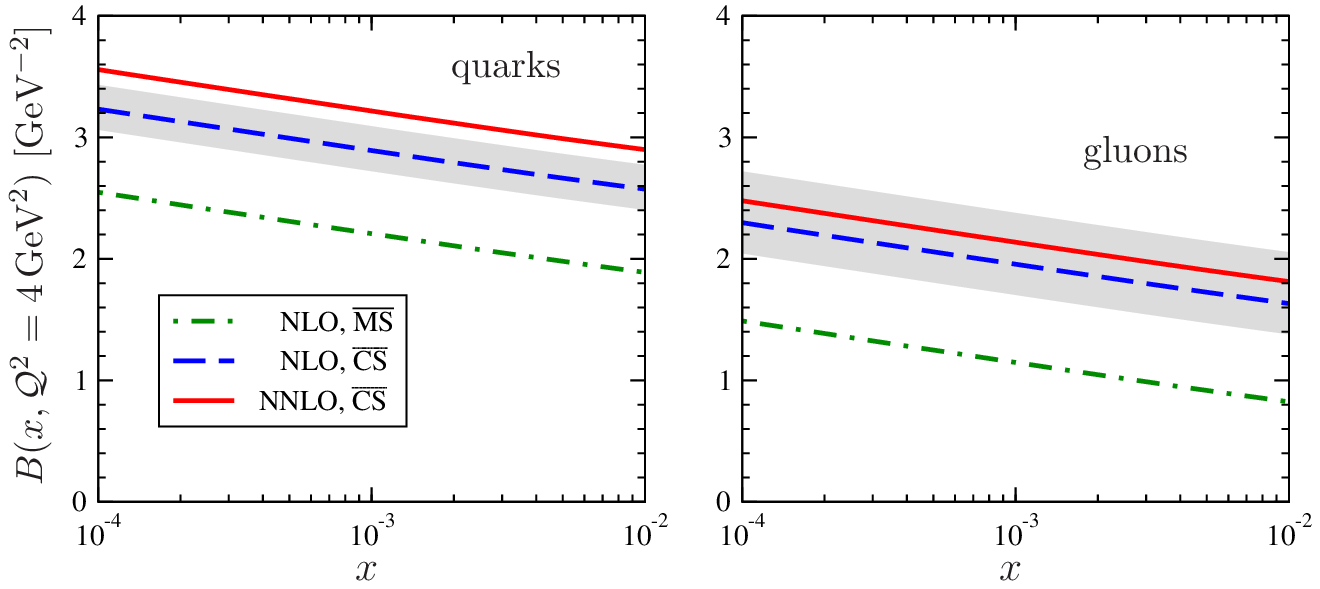}%
&
\includegraphics[scale=0.52]{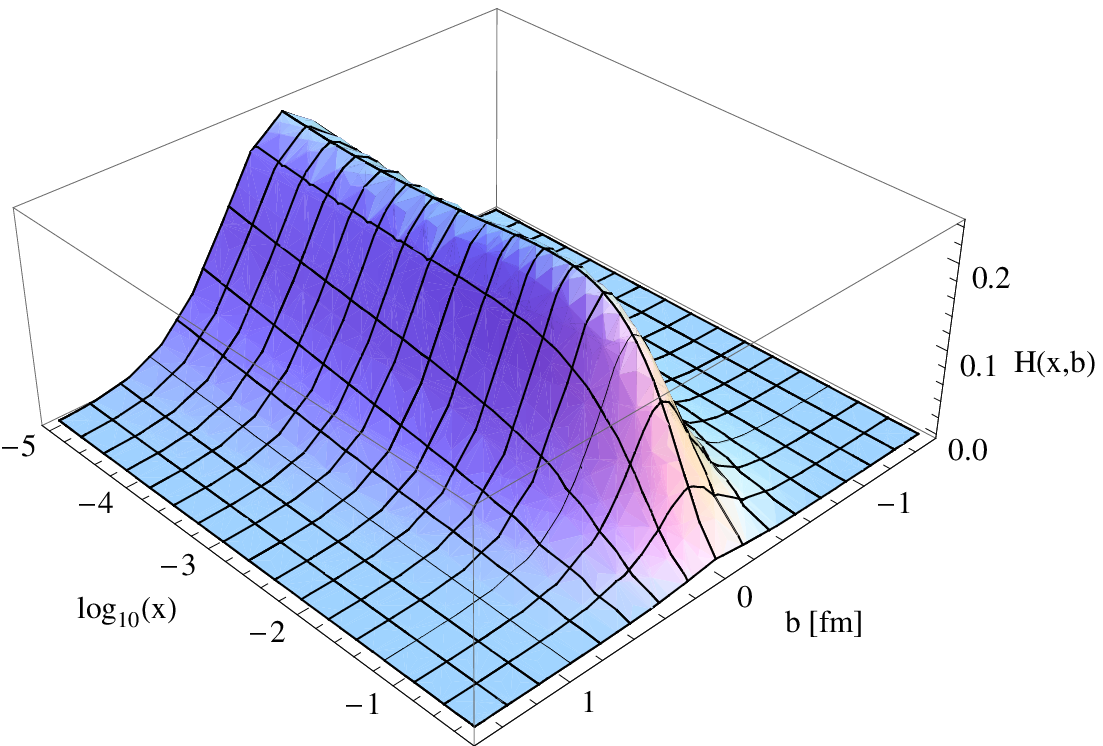}
\\
a) & b)
\end{tabular}
\vspace{-0.3cm}
\caption{a) Resulting GPD slope $B=\langle {\bf b}^2 \rangle / 4$
at the input scale ${\cal Q}^2 = 4\,{\rm GeV}^2$ and
b) 3D picture of gluon GPD
(\protect\ref{Def-Hq-eta0}). }
\label{f:B}
\end{figure}

Turning now to presentation of the GPDs resulting from the fits,
we recall that Fourier transform of GPDs for $\eta=0$,
\begin{eqnarray}
\label{Def-Hq-eta0}
H(x,{\bf b}) =
\int\!\frac{d^2{\bf \Delta}}{(2\pi)^2}\,
e^{-i {\bf b}\cdot {\bf \Delta}} H(x,\eta=0,\Delta^2=-{\bf \Delta}^2)\,,
\end{eqnarray}
can be interpreted in the infinite momentum frame as probability
density \cite{Burkardt00,*Burkardt02}, see Fig.~\ref{f:B}b. The
average transversal parton distance squared $\langle {\bf b}^2
\rangle$ is  given by the GPD slope $B=\langle {\bf b}^2 \rangle
/4$, shown in Fig.~\ref{f:B}a. Although gluons are perturbatively
suppressed in DVCS, one has a handle on them via the ${\cal Q}^2$
evolution. The found value for $\langle {\bf b}^2 \rangle$ is
compatible with the analysis for deeply exclusive photo and
leptoproduction of $J/\psi$ \cite{StrWei03}. The results confirm
the picture, mentioned in the introduction, about the correlation
of transversal and longitudinal degrees of freedom: partons with
smaller momentum fractions are more decentralized. Intuitively,
this  small $x$ partons might be somehow associated with the meson
cloud that surrounds the proton center.

\section{Summary}

GPDs encode a unified description of the proton structure  and
they are experimentally accessible via the hard leptoproduction of
photon or mesons. We have shown that the representation of Compton
form factors as Mellin-Barnes integrals offers a useful tool in
analyzing DVCS: the inclusion of evolution is simple, numerical
treatment is stable and fast. Although the motivation for this
representation originated from manifest conformal symmetry at LO,
we have shown here that this Mellin--Barnes integral
representation can be used within the standard $\overline{\rm MS}$
scheme beyond LO. Such a representation can be straightforwardly
obtained from the momentum fraction representation and, therefore,
also other GPD related processes, e.g., the hard electroproduction
of mesons, can be given in terms of Mellin--Barnes integrals. This
opens a new road for the `global' analysis of experimental data
within the perturbative GPD formalism to NLO accuracy.
Furthermore, the use of conformal symmetry enables elegant
approach to higher-order radiative corrections to the DVCS
amplitude. We have shown that although NLO corrections can be
sizable, and are strongly dependent on the gluonic input, the NNLO
corrections are small to moderate, supporting perturbative
framework of DVCS. The observed change in the scale dependence is
not so conclusive: similarly as in DIS we encounter large NNLO
effects for $\xi < 10^{-3}$, which signal a breakdown of naive
perturbation series expansion of the evolution operator.
Nevertheless, this breakdown is universal and if we precisely
define the treatment of this operator, perturbative QCD can be
employed as a tool for analyzing data even in the small $\xi$
region. We expect that within increasing accuracy in the
perturbative approximation the GPD formalism improves.

Finally, fits to available DVCS and DIS data in collider
kinematics work well in the collinear factorization approach
beyond LO and give access to transversal distribution of partons.
The lesson learned from the failure of LO fits can be extended to
(oversimplified) GPD models used for phenomenology in fixed target
kinematics. They do not possess a flexible skewness dependence
parameterization and so the normalization of the amplitude
for fixed $t$ is mostly determined. Ironically, this non-flexibility
originates from the implementation of PDF and form factor
constraints in the most convenient manner. This can be repaired in
the Mellin-Barnes representation by a model dependent resummation
of SO(3) partial waves or in the momentum fraction representation
by a more flexible double distribution ansatz, in which also the
correlation of $t$- and skewness dependence is improved
\cite{HwaMue07}.

Additionally, we have observed that the off-shell Regge-exchange
amplitudes\cite{SzcLonLla07}, having the generic $s$-dependence of
high-energy on-shell amplitudes, are ruled out. Repairing this
failure yields the well-known dilemma of Regge-phenomenology.
Namely, when extended to off-shell processes it fails to describe
high energy data and must be improved, thereby loosing its
predictive power \cite{DonLan98,*CapFazFioJenPac06}. This loss of
predictive power appears also in description of deeply exclusive
leptoproduction data in the valence quark region within Regge
inspired {\em models}, often outside of the accepted validity
region $\{25\, {\rm GeV}^2 \lesssim W^2, -t \lesssim 1 {\rm
GeV}^2\}$ in classical Regge phenomenology, see
e.g.~\cite{CanLag02}. Certainly, these approaches offer an easy
and economical possibility to fit data and to interpret them in
terms of mesonic exchanges and numerous couplings (unknown form
factors). In our opinion the GPD formalism offers a new
perspective for our QCD understanding, including a handle on the
dynamical sources of Regge behavior. In this formalism one
extracts information about non-perturbative quantities from
experimental observables and interprets it in terms of partonic
degrees of freedom. We consider Regge-inspired interpretations of
data \cite{Lag07} as valuable tool for building realistic GPD
models. This gives one handle more on the challenging task to
understand hadron physics from QCD dynamics. Non-perturbative GPD
aspects, in particular Regge behavior, might be studied within
(partial) resummation of $t$-channel ladders (in the language of
light-cone wave functions of higher partonic Fock states), e.g.,
Ref.~\cite{BalKuc00,ErmGreTro00}, lattice simulations, and model
building. This might lead to clues to improved understanding of
hadron physics in terms of the underlying theory.

\vspace{3mm}
\pagebreak[2]
\noindent
{\bf Acknowledgements}

\noindent
For illuminating discussions on  Regge, BFKL, and collinear
factorization approaches and their interplay  as well as on GPD
modelling we like to thank S.~Brodsky, M.~Diehl, B.~Ermolaev,
V.~Fadin, D.~Ivanov, L.L.~Jenkovszky, P.~Kroll, M.~Polyakov, and
A.~Radyushkin. In particular, we are grateful to A.~Szczepaniak for
discussions on the work presented in Ref.~\cite{SzcLonLla07}.
D.M.~is indebted to A.~Miller, A.V.~Efremov and O.~Teryaev for
invitation and support, allowing him to participate on {\it The
6th~Circum-Pan-Pacific Symposium on High Energy Spin Physics} and
the {\it XII  Workshop on High Energy Spin Physics}, respectively.
K. P-K. would like to thank the organizers of the
{\it 12th International Conference
on Elastic and Diffractive Scattering}, DESY Hamburg,
for invitation and support.

\begin{footnotesize}
\providecommand{\etal}{et al.\xspace}
\providecommand{\href}[2]{#2}
\providecommand{\coll}{Coll.}
\catcode`\@=11
\def\@bibitem#1{%
\ifmc@bstsupport
  \mc@iftail{#1}%
    {;\newline\ignorespaces}%
    {\ifmc@first\else.\fi\orig@bibitem{#1}}
  \mc@firstfalse
\else
  \mc@iftail{#1}%
    {\ignorespaces}%
    {\orig@bibitem{#1}}%
\fi}%
\catcode`\@=12

\end{footnotesize}
\end{document}